# WALLABY Early Science - IV. ASKAP HI imaging of the nearby galaxy IC 5201

D. Kleiner[1,2]⋆, B. S. Koribalski[2], P. Serra[1], M. T. Whiting[2], T. Westmeier[3], O. I. Wong[3,4], P. Kamphuis[5], A. Popping[3], G. Bekiaris[2], A. Elagali[3,2,4], B.-Q. For[3,4], K. Lee-Waddell[2], J. P. Madrid[2], T.N. Reynolds[3,2,4], J. Rhee[3,4], L. Shao[6,7], L. Staveley-Smith[3,4], J. Wang[6], C. S. Anderson[8], J. Collier[2,9,10], S. M. Ord[2], M. A. Voronkov[2]

[1] *INAF – Osservatorio Astronomico di Cagliari, Via della Scienza 5, 09047 Selargius (CA), Italy*
[2] *CSIRO Astronomy & Space Science, Australia Telescope National Facility, PO Box 76, Epping, NSW 1710, Australia*
[3] *International Centre for Radio Astronomy Research (ICRAR), The University of Western Australia, 35 Stirling Hwy, Crawley, WA, 6009, Australia*
[4] *ARC Centre of Excellence for All Sky Astrophysics in 3 Dimensions (ASTRO 3D)*
[5] *Astronomisches Institut der Ruhr-Universität Bochum (AIRUB), Universitätsstrasse 150, D-44801 Bochum, Germany*
[6] *Kavli Institute for Astronomy and Astrophysics, Peking University, Beijing 100871, China*
[7] *Research School of Astronomy and Astrophysics, Australian National University, Canberra ACT 2611, Australia*
[8] *CSIRO Astronomy & Space Science, 26 Dick Perry Avenue, Kensington, WA 6151, Australia*
[9] *School of Computing, Engineering and Mathematics, Western Sydney University, Locked Bag 1797, Penrith, NSW 2751, Australia*
[10] *The Inter-University Institute for Data Intensive Astronomy (IDIA), Department of Astronomy, University of Cape Town, Rondebosch, 7701, South Africa*

**ABSTRACT**

We present a Wide-field ASKAP L-Band Legacy All-sky Blind surveY (WALLABY) study of the nearby ($v_{sys} = 915$ km s$^{-1}$) spiral galaxy IC 5201 using the Australian Square Kilometre Array Pathfinder (ASKAP). IC 5201 is a blue, barred spiral galaxy that follows the known scaling relations between stellar mass, SFR, H I mass and diameter. We create a four-beam mosaicked H I image cube, from 175 hours of observations made with a 12-antenna sub-array. The RMS noise level of the cube is 1.7 mJy beam$^{-1}$ per channel, equivalent to a column density of $N_{HI} = 1.4 \times 10^{20}$ cm$^{-2}$ over 25 km s$^{-1}$. We report 9 extragalactic H I detections – 5 new H I detections including the first velocity measurements for 2 galaxies. These sources are IC 5201, 3 dwarf satellite galaxies, 2 galaxies and a tidal feature belonging to the NGC 7232/3 triplet and 2 potential infalling galaxies to the triplet. There is evidence of a previous tidal interaction between IC 5201 and the irregular satellite AM 2220−460. A close fly-by is likely responsible for the asymmetric optical morphology of IC 5201 and warping its disc, resulting in the irregular morphology of AM 2220−460. We quantify the H I kinematics of IC 5201, presenting its rotation curve as well as showing that the warp starts at 14 kpc along the major axis, increasing as a function of radius with a maximum difference in position angle of 20°. There is no evidence of stripped H I, triggered or quenched star formation in the system as measured using DECam optical and *GALEX* UV photometry.

**Key words:** galaxies: evolution – galaxies: general – galaxies: groups: individual: IC 5201 – galaxies: interactions – radio lines: galaxies

## 1 INTRODUCTION

To understand how galaxies form and evolve, it is crucial to understand the gas content of the galaxies. A combination of nature and nurture drives the evolution of galaxies (Casado et al. 2015) and there are many internal and environmental mechanisms that work in tandem to transform and evolve galaxies (e.g. Gunn & Gott 1972; Toomre & Toomre 1972; Dressler 1980; Butcher & Oemler 1984; Moore et al. 1996; Lewis et al. 2002; Gómez et al. 2003; Baldry et al. 2004; Bell et al. 2004; Balogh et al. 2004; Schawinski et al. 2014; Davies et al. 2019). See Conselice (2014) and references within for a comprehensive review of the transformation mechanisms affecting galaxy evolution. The gas in galaxies is more easily disrupted compared to the stellar component, making it sensitive to internal and external physical processes. In particular, observing neutral hydrogen in its atomic form (H I) with radio telescopes is used to quantify the cool gas content.

⋆ E-mail: Dane.Kleiner@inaf.it

H I is the dominant component of the interstellar medium (ISM) in late type galaxies and is ideal for tracing tidal interactions and environmental processes (Doyle & Drinkwater 2006; Kilborn et al. 2009; Cortese et al. 2011; Catinella et al. 2010, 2013; Lee-Waddell et al. 2016; Hess et al. 2019). The H I often extends well beyond the bright stellar component (e.g. Koribalski et al. 2018), making it useful for probing the dark matter (DM) content (e.g. Warren et al. 2004), as well as the kinematics, dynamical history (e.g. Yun et al. 1994; Koribalski & Dickey 2004; Kurapati et al. 2018), local and large scale accretion processes (e.g. de Blok et al. 2018; Kleiner et al. 2017). H I is fundamental to galaxy evolution and a galaxy with no H I is going to cease forming stars.

Widefield H I surveys have typically been conducted in the past with large single dishes. The large collecting area of a single dish results in a good sensitivity being reached with short integration times and multibeam receivers (e.g Staveley-Smith et al. 1996) enable fast sky surveys. Therefore, large single dishes are useful for observing a wide area of the sky in a reasonable amount of time. However, single dishes have coarse spatial resolution, often on the order of several arcminutes, that are often unable to resolve the H I in nearby galaxies. Two prime examples are the H I Parkes All Sky Survey (HIPASS; Barnes et al. 2001) and the Arecibo Legacy Fast ALFA (ALFALFA; Giovanelli et al. 2005) survey, that have catalogued and measured the properties (e.g. H I mass function) of thousands of H I galaxies in the nearby Universe (e.g. Meyer et al. 2004; Koribalski et al. 2004; Zwaan et al. 2005; Jones et al. 2018; Haynes et al. 2018).

The Square Kilometre Array (SKA) pathfinder and precursor telescopes are interferometers that are currently being commissioned. The telescopes sensitive to nearby H I are: The Australian Square Kilometre Array Pathfinder (ASKAP; Johnston et al. 2007; DeBoer et al. 2009), the Meer Karoo Array Telescope (MeerKAT; Booth et al. 2009), the APERture Tile In Focus (APERTIF; Verheijen et al. 2008) and the upgraded Giant Metrewave Radio Telescope (uGMRT; Gupta et al. 2017). Each of these instruments have an improved spatial resolution by $1 - 2$ orders of magnitude compared to single dish telescopes. MeerKAT has a sensitivity that rivals most of the biggest single dish telescopes, ASKAP and APERTIF have instantaneous wide-field imaging ability. While each telescope has its niche (ASKAP is discussed in the next section), they have the ability to survey the H I sky at unprecedented speeds with improved resolution and sensitivity.

In this paper, we use ASKAP during its early science and commissioning phase to image the nearby barred spiral galaxy IC 5201, as part of the Wide-field ASKAP L-Band Legacy All-sky Blind surveY (WALLABY; Koribalski 2012) early science. Using the new H I images, we measure the kinematics and use ancillary multiwavelength data to quantify the stellar component and star formation properties. This work is necessary for (and contributed to) verifying the observing technique of ASKAP with its new receivers, as well as how the ASKAP data reduction pipeline images bright, well-resolved H I galaxies. These early science results, including those from Reynolds et al. (2019), Lee-Waddell et al. (2019) and Elagali et al. (2019), show the variety of results that will be produced by WALLABY.

This paper is structured in the following way; Section 2 describes the ASKAP observations and pertinent details of WALLABY early science. In Section 3 we describe the galaxy IC 5201 and why it was chosen for imaging. The optical and UV ancillary data is presented in Section 4, with the data reduction processes of all observations described in Section 5, presenting the relevant and important details of ASKAPsoft, required for early science. The source finding and analysis of the data is described in Section 6. We present the results of our H I measurements, images, kinematics, stellar and star formation properties in Section 7, discussing the implications and likely scenario for our findings in Section 8. Finally, we present our conclusions in Section 9. A standard, flat, $\Lambda$ cold dark matter cosmology has been assumed throughout this paper, using $\Omega_M = 0.3$, $\Omega_\Lambda = 0.7$ and $H_0 = 70\ \mathrm{km\,s^{-1}\ Mpc^{-1}}$ (Hinshaw et al. 2013).

## 2   ASKAP OBSERVATIONS AND WALLABY EARLY SCIENCE

ASKAP is a new radio interferometer consisting of $36 \times 12$-m antennas, each equipped with a second-generation (Mk II) Phased Array Feed (PAF). ASKAP is a powerful survey telescope due to the instantaneous (30 deg$^2$ at $\sim 1.4$ GHz) wide field of view from the PAFs and 300 MHz bandwidth (Johnston et al. 2007; DeBoer et al. 2009). The PAFs were configured to observe the 36 beams in a $6 \times 6$ square footprint (Fig. 1) for these observations.

Our target galaxy, IC 5201, is located in the foreground of the NGC 7232 group (Fig. 1), which was observed as the first WALLABY Early Science field. Details of the observations are given in Reynolds et al. (2019, Paper I), who focused on the NGC 7162 galaxy group, and Lee-Waddell et al. (2019, Paper II), who studied the NGC 7232 triplet and surroundings[1]. The details of WALLABY early science and the NGC 7232 observations can be found in the aforementioned papers, though we inform the reader of the relevant information for this work:

---

[1] There is a small region of overlap between the images from this work and the images in Paper II. We utilise this overlap to independently test and verify the H I flux and velocity of suitable detections. The RMS noise of the overlap in Paper II is $\geq 30\%$ lower and sampled more uniformly than our image, and we therefore refer the reader to Paper II for the detailed analysis of this region.

- Between $10 - 12$ antennas were used for these early science observations (Table. 1), spending a total of 175 hours on source – sufficient time to reach the expected WALLABY Root Mean Square (RMS) noise level of $\sim 1.7$ mJy beam$^{-1}$ per 4 km s$^{-1}$ channel, with a resolution of $\sim 30$ arcsec (Duffy et al. 2012; Koribalski 2012).

- The observations consist of two (A & B) interleaved pointings, in order to achieve uniform sensitivity across the field (Fig. 1). An additional pointing (footprint B0) was also used in this work, it has the same central pointing as Footprint B, with a position angle of $0°$ (Table. 1).

- We extract a 12 MHz (1408.5 - 1420.5 MHz) sub-band from beams 9 and 10 in footprint A along with beams 10 and 25 from footprint B (Fig. 1) to create a mosaic with IC 5201 in the centre.

- We present the details of the individual ASKAP observations in Table 1. Two of the early commissioning observations are excluded from this analysis as the measured fluxes do not agree with HIPASS catalogued fluxes.

## 3 THE BARRED SPIRAL GALAXY IC 5201

IC 5201 is a nearby, gas-rich spiral galaxy of a type SB(rs)cd (de Vaucouleurs et al. 1991), with a Hubble flow distance of 13.2 Mpc and systemic velocity of 915 km s$^{-1}$. It is the brightest H I source in the NGC 7232 field with a catalogued integrated H I flux of $F_{HI} = 165.5 \pm 12.0$ Jy km s$^{-1}$, $M_{HI} = 10^{9.74}$ M$_\odot$ and 20% velocity width of $w_{20} = 209$ km s$^{-1}$ as measured in the HIPASS Bright Galaxy Catalogue (Koribalski et al. 2004). The photometric properties of IC 5201 are shown in Table 2.

IC 5201 is a relatively isolated galaxy (Fig. 1) and it appears to be a typical spiral that follows the known scaling relations, such as; the H I-to-stellar mass scaling relation for H I detections (Brown et al. 2015) and the galaxy has a high specific star formation rate (SSFR), consistent for a blue, spiral galaxy with a stellar mass of $\sim 10^{9.8}$ M$_\odot$ (Meurer et al. 2006; Saintonge et al. 2016). However, the spiral arms of IC 5201 are asymmetric (Fig. 1), which may suggest the galaxy has undergone non-secular evolution, such as interactions with other galaxies.

The bright H I flux of IC 5201 makes it ideal for tracing each step in the data reduction, specifically the flux and noise properties of the spectral line images. It is also a good test of the mosaicking as IC 5201 is located at the intersection of multiple beams in footprint B and between beams of footprints A and B (Fig. 1).

However, given that IC 5201 is a bright extended H I source, it is difficult to clean properly and image without artefacts (see section 5.1.4 for details on how this is done) with the ASKAP 12-antenna array. If the images are poorly cleaned, source finding will be severely hindered and faint nearby galaxies will be difficult, if not impossible to detect.

In comparison to the HIPASS images, the ASKAP images of IC 5201 allows resolved mapping of the galaxy and its immediate environment. Additionally, the images and spectra produced in this work will have an improved RMS, velocity and angular resolution by factors of $\sim 8$, 4.5 and 30. These improvements are applicable to the $\sim 5000$ nearby bright H I sources that ASKAP will image, previously found by HIPASS (Meyer et al. 2004; Koribalski et al. 2004; Wong et al. 2006).

## 4 ANCILLARY DATA

### 4.1 DECam - optical data

We make use of the optical/near-infrared images that overlap with our ASKAP images from the Dark Energy Survey (DES; Abbott et al. 2018) DR1. DES is a 5000 square degree survey of the Southern sky in the $g, r, i, z$ and $Y$ optical bands with the Dark Energy Camera (DECam; Flaugher et al. 2015) mounted on the 4-m Blanco telescope at Cerro Tololo Inter-American Observatory.

DES DR1 consists of images and source catalogues assembled over the first three years of DES science operations. The photometric precision is < 1% in all bands, and the astrometric precision is 151 mas. The median coadded depth is $g = 24.33$, $r = 24.08$, $i = 23.44$, $z = 22.69$, and $Y = 21.44$ mag for a 1.95" diameter aperture with a S/N of 10 (Abbott et al. 2018).

### 4.2 *GALEX* - UV data

We use archival far- and near-ultraviolet (FUV and NUV) observations with pivot wavelengths at 1524-Å and 2273-Å from the Galaxy Evolution Explorer (GALEX) satellite telescope (Morrissey et al. 2005; Bianchi 2014). The *GALEX* observations typically have a spatial resolution of approximately 5–6″ (Wyder et al. 2005). Shallow All-sky Imaging Survey (AIS) *GALEX* observations are obtained for the entire IC 5201 group presented in this paper. The stellar and star formation properties of IC 5201 were measured and discussed as part of the Survey of Ultraviolet emission of Neutral Gas Galaxies (SUNGG; Wong et al. 2016).

## 5 DATA REDUCTION

### 5.1 The ASKAPsoft pipeline

ASKAPsoft[2] (Whiting et al. in prep) is the custom data reduction software used to calibrate ASKAP observations and produce science images and catalogues. The ASKAPsoft modules are specifically designed to run on supercomputers in order to handle the large data rates and volumes produced by ASKAP. The ASKAPsoft modules are wrapped into a pipeline that enabled the automated processing of ASKAP observations. In this work, ASKAPsoft versions $0.18.3 - 0.19.7$ were used to reduce the data. The visibilities obtained in each observation are directly sent to the Pawsey Supercomputing Centre, where an ingest pipeline creates measurement sets after flagging data affected by hardware (e.g. antenna, correlator) issues. During early science, the raw visibilities were stored on disk to enable thorough testing that resulted in significant improvements to ASKAPsoft and the pipeline. This is in contrast to the anticipated situation for the 36-antenna array, where the data rates and volumes will be so large that it is impractical to

---

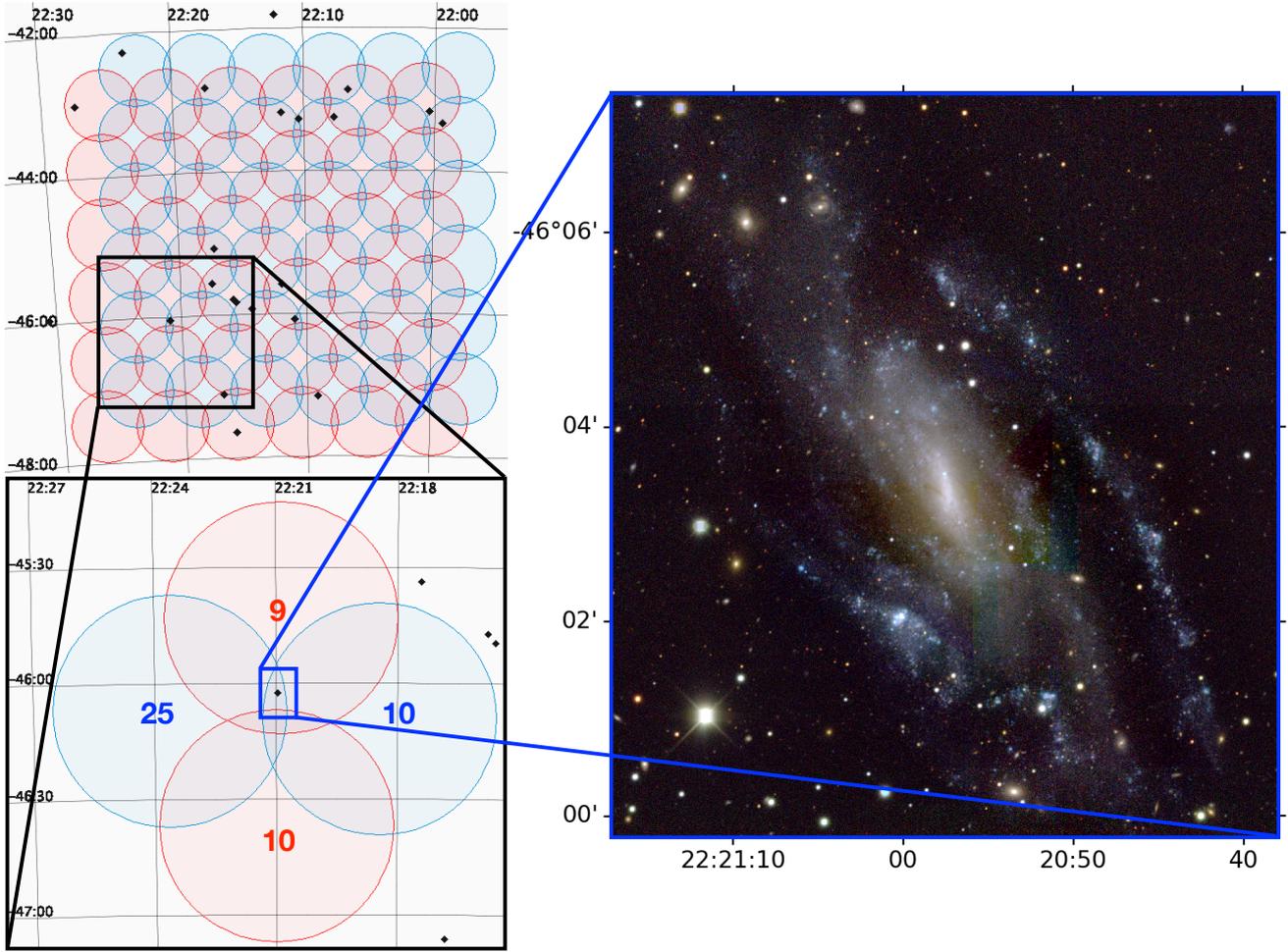

**Figure 1.** Top-left: ASKAP 36-beam interleaved footprints A (red) and B (blue) of the NGC 7232 field with catalogued HIPASS sources as black diamonds. The bottom-left (black) insert shows the four ASKAP beams and beam numbers used in this analysis along with the HIPASS sources in the field. The right (blue) insert shows the optical 3-colour (*gri*) composite of our main target, the barred spiral galaxy IC 5201, observed with the Dark Energy Camera as part of the Dark Energy Survey (Abbott et al. 2018). The axes show the J2000 Right Ascension and Declination.

store the raw and intermediate data products for lengthy periods.

For early science, `ASKAPsoft` has been run and customised by the users. The following sections describe the core data reduction and imaging processes used in all of the WALLABY early science papers (Reynolds et al. 2019; Lee-Waddell et al. 2019; Elagali et al. 2019), where any small deviation is described in their respective paper. The `ASKAP-soft` pipeline operates on an individual SBID (e.g. Table 1) at a time, and is responsible for the calibration, flagging and imaging. Fig. 2 shows the default (blue) `ASKAPsoft` pipeline and the custom (red) method we have used in this work. The default `ASKAPsoft` pipeline was run on each dataset up until the point of the continuum validation. This produced a wide-field continuum image with measured validation metrics on the image. We then re-reduce the beams containing IC 5201 in each dataset by processing them with `ASKAPsoft` a second time, correcting any astrometric offsets from the component positions used in the self-calibration (Fig. 2). After the pipeline produced calibrated, position-corrected visibilities, we used a custom method to image and mosaic the data. Be-low, we outline and describe the main features of `ASKAPsoft`, including any deviation from the default pipeline.

### 5.1.1 Bandpass calibration and flagging

Each data set is calibrated separately on a per-beam basis for each observation. After the removal of radio frequency interference (RFI) from the primary flux calibrator PKS 1934–638, it is first used to calibrate both the flux scale and bandpass shape for each beam. Then dynamic amplitude flagging is applied to remove outliers from the science visibilities. This includes channels affected by (known, suspected and detected) RFI and bad beams, antennas or baselines. The science visibilities were copied and averaged into 1 MHz wide channels that underwent a second round of dynamic flagging, removing RFI that did not exceed the flagging threshold applied to an individual fine channel.

**Table 1.** A summary of the WALLABY Early Science observations of the NGC 7232 field, including the Scheduling Block ID (SBID), a unique identifier that can be searched and download the data at the CSIRO ASKAP Science Data Archive (CASDA; https://casda.csiro.au/). There are two main interleaving footprints (A & B), however, one observation of the B footprint was not rotated. We denote this as B0 and is imaged separately to the other footprint B observations, that we describe in further detail in section 5.1.4. Two early commissioning observations (~ 20 hrs) are excluded from this Table and analysis, due to calibration uncertainties.

| Date (2016) | SBID | Number of Antennas | Time on-source (Hr) | Footprint | Position Angle (°) |
|---|---|---|---|---|---|
| August 11 | 1927 | 10 | 12.0 | A | 0 |
| August 12 | 1934 | 11 | 11.5 | A | 0 |
| October 7 | 2238 | 11 | 10.4 | B0 | 0 |
| October 8 | 2247 | 12 | 11.4 | B | -0.4 |
| October 9 | 2253 | 12 | 12.1 | B | -0.4 |
| October 10 | 2264 | 12 | 11.0 | B | -0.4 |
| October 11 | 2270 | 12 | 12.5 | A | 0 |
| October 12 | 2280 | 12 | 9.2 | B | -0.4 |
| October 13 | 2289 | 10 | 11.3 | A | 0 |
| October 14 | 2299 | 12 | 10.4 | B | -0.4 |
| October 16 | 2325 | 12 | 12.4 | A | 0 |
| October 17 | 2329 | 12 | 12.3 | B | -0.4 |
| October 18 | 2338 | 12 | 12.0 | A | 0 |
| October 19 | 2347 | 12 | 12.0 | B | -0.4 |
| Date (2017) | | | | | |
| August 23 | 4002 | 11 | 10.0 | A | 0 |
| September 27 | 4399 | 12 | 5.0 | A | 0 |

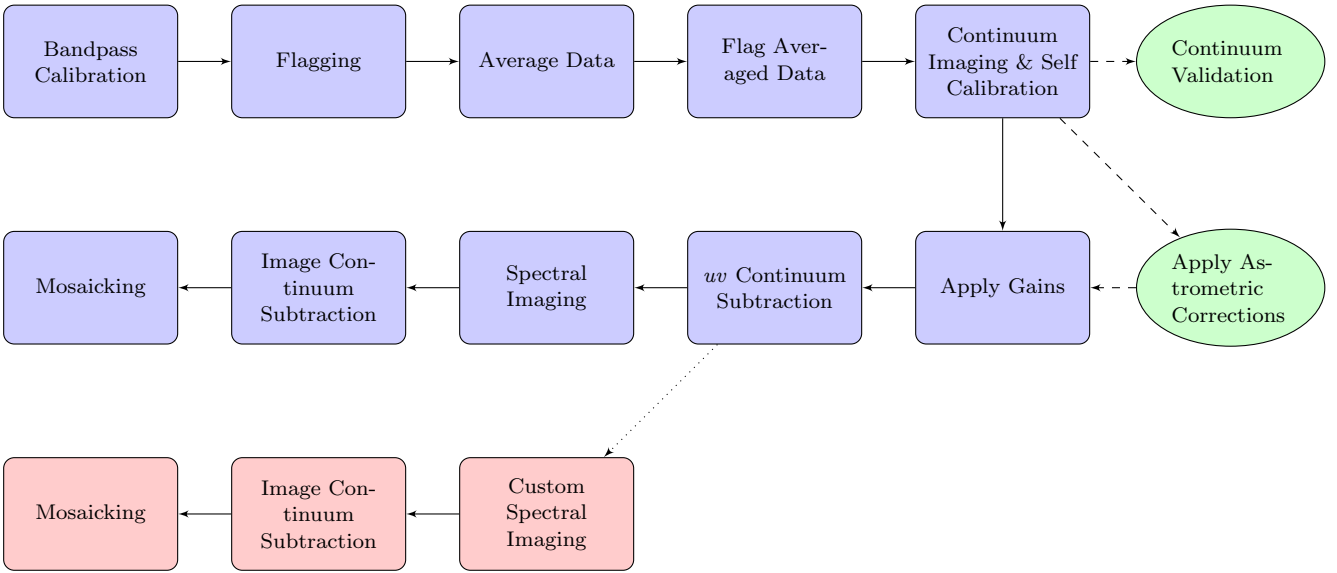

**Figure 2.** Flow chart outlining the main steps of the `ASKAPsoft` pipeline. The blue blocks show the default steps in the `ASKAPsoft` pipeline, the green ovals show processes than can be included in the pipeline and the red blocks shows the custom method used to produce the spectral images in this work. The default `ASKAPsoft` pipeline was first used on all observations up to the point of the continuum validation. This produced a wide-field 36 beam continuum image for each data set with astrometric offsets. The default pipeline was then run a second time on each dataset up to the point of the *uv*-based continuum subtraction, producing the astrometrically corrected, calibrated visibilities in preparation for the spectral imaging. We then deviate from the default pipeline to do custom spectral imaging on each individual beam. The custom spectral imaging uses the core `ASKAPsoft` modules, but in a different technique that is not implemented in the pipeline. See section 5.1.4 and beyond for the details of the imaging technique.

### 5.1.2 Continuum imaging and self-calibration

The calibrated, flagged and averaged science visibilities are processed to create continuum images for each beam. Time-varying phase errors in the continuum images are minimised through the technique of self-calibration. This is achieved by running the Selavy source finder (Whiting & Humphreys 2012) on the continuum images to produce a catalogue. A model is created from the sources in the catalogue and is used to calibrate the time-dependent complex gains. As the continuum image is improved through updating the gains, a further two self-calibration loops are applied to the continuum images, repeating the process of creating a model and refining the gains.

In our first run of ASKAPsoft, the 36 individual continuum images from each beam are mosaicked and the vali-dation of the wide-field image is assessed by measuring the flux, spectral index and position of each source compared to the sources in the Sydney University Molonglo Sky Survey (SUMSS – Mauch et al. 2003). At the time of these observations, there were small astrometric offsets[3], typically on the order of ~ 4 arcsec, that were measured from the wide-field continuum images. The absolute astrometry of the ASKAP continuum images are boot-strapped from the SUMSS source positions.

**Table 2.** The photometric properties of IC 5201 compiled from the literature. The photometry is measured in the Carnegie-Irvine Galaxy Survey (Ho et al. 2011; Li et al. 2011), where the *b*-band magnitude is the total (*bt*) apparent magnitude from HyperLeda, the position angle is derived from the *b*-band major axis and the diameter is measured at $\mu_B = 26.5$ mag arcsec$^{-2}$ while being corrected for Galactic extinction and inclination.

| Quantity | Value | Reference |
|---|---|---|
| J2000 $\alpha$ (hms) | 22 20 57.39 | Li et al. (2011) |
| J2000 $\delta$ (° ′ ″) | -46 02 09.7 | Li et al. (2011) |
| Distance (Mpc) | 13.2 | This work |
| Hubble (RC3) Type | SB(rs)cd | de Vaucouleurs et al. (1991) |
| *b*-Band (mag) | 11.95 | Li et al. (2011) |
| Position Angle (°) | 25 ± 6 | Li et al. (2011) |
| Axis Ratio | 1.83 | Lauberts & Valentijn (1989) |
| Inclination (°) | 64.8 ± 6 | Li et al. (2011) |
| Diameter (′) | 8.85 | Li et al. (2011) |

In our second run of ASKAPSOFT, the bulk astrometric corrections are applied in the last self-calibration loop to ensure that all final continuum images and calibration tables have the correct positions. The calibration tables (including the time-dependent gains solution) are then applied to the non-averaged (18.5 kHz channel width), flagged science visibilities in preparation for spectral line processing.

### 5.1.3  *uv-based continuum subtraction*

Before imaging the large volume of H I spectral line data, ideally all radio continuum emission needs to be subtracted from the calibrated visibilities. For ASKAP data this is done in two steps, the first step is to subtract the continuum emission in the visibilty domain. First, a model of the continuum image is constructed using the SELAVY task, that fits 2D Gaussian components to each source detected in the image. These components are then used to create a deconvolved sky model image that simulate visibilities. the model is then subtracted from the observed visibilities, creating a new dataset which can then be imaged.

### 5.1.4  *HI imaging*

The spectral imaging within ASKAPsoft images each channel separately, creating a spectral cube with the same spectral resolution as the input measurement set. The visibilities for each channel are gridded in the *uv*-domain, inverted, then deconvolved and restored to produce the final cube.

The default mode of the ASKAPsoft pipeline is to operate on a single dataset at once. For this work, however, we provide the ASKAPsoft imager module with multiple datasets that are jointly gridded prior to the Fourier transform. This approach requires, for now, all datasets to have the same pointing centre - that is, they must be from the same beam. This imaging is done through a custom method run outside the standard ASKAPsoft pipeline.

The greatest advantage of imaging a total of $N$ measurement sets of calibrated visibilities on the same *uv* grid is that the deconvolution cleans deeper by a factor of $\sqrt{N}$. Therefore, for an individual beam in a given footprint, we provide all of the calibrated, continuum subtracted visibilities to be imaged. This equates to two images per beam in each footprint, i.e. 8 nights in footprint A, 7 nights in footprint B and 1 night in footprint B0.

As IC 5201 is an extended, bright H I source, it is difficult to clean and will produce bright sidelobes unless extra care is taken in the imaging process. We have extensively tested different imaging techniques that focused on the weighting and cleaning parameters. Our final images have suppressed sidelobe emission to less than $1\sigma$ of the noise. This was achieved by utilising two features implemented in the ASKAPsoft imager: multiscale clean and deep clean.

Multiscale clean uses clean components of different sizes to the emission in the spectral cube (Rau & Cornwell 2011). The advantage of this over a traditional algorithm such as Högbom is that it can better model the complex, extended emission using different sized components (Rich et al. 2008). We found that a combination of four components of deconvolved pixel sizes [0,3,10,30], corresponding to point sources, small, medium and large-scale emission was the most effective in cleaning the channels containing emission from IC 5201.

The ASKAPsoft imager has the ability to clean spectral images in the *uv*-plane and in the image-plane, i.e. major and minor cycles. The *uv*-plane cleaning occurs through removing the clean component image (clean model) from the visibilities and gridding/degridding during the major cycle is achieved using the WProjection (w-snapshots in this instance) algorithm to account for wide-field aberration (Cornwell et al. 2012). The image-based cleaning removes emission peaks in the image-plane (equivalent to deconvolution), that in-turn is used to build and improve the clean model. We also utilise a deep clean (analogous to masking), that searches for peaks down to a lower ($0.5\sigma$ of the expected RMS) threshold, but only in the pixels already in the model i.e. the known clean components.

We create spectral cubes for the beam 9 and 10 in footprint A and 10 and 25 in footprint B and B0. Our cubes contain 650 channels and were created with a weighting of robust = 0.5 and no taper. The channel width is ~ 4 km s$^{-1}$ and covers a barycentric velocity range of –20 to 2513 km s$^{-1}$ (1408.5 - 1420.5 MHz).

### 5.1.5  *Image-based continuum subtraction*

To obtain high quality H I data cubes, we find it necessary to subtract residual continuum emission in the image domain. The bright continuum sources can result in a residual at a level of 1–3% of the initial flux density in the final H I cube. For example, the brightest source in a beam, ini-tially has a peak flux density of 320 mJy beam$^{-1}$. After the first *uv*-continuum subtraction, the peak flux density of the source is 49 mJy beam$^{-1}$ and after the image-based continuum subtraction, the final flux density is 6 mJy beam$^{-1}$. Continuum residuals will adversely affect H I source finding and parameterisation, though we are able to easily identify bad sightlines caused by continuum residuals in the H I cube (described in Section 6).

---

[3] Due to inadequate phasing model of the array resulting in a slight inaccuracy of the phase centre for the ASKAP footprint. This issue is specific to early science data and was resolved in November 2017.

The continuum sky model, which is subtracted from the ASKAP visibilities, will much improve as the array grows towards the full 36-antenna interferometer, due to improved *uv*-coverage and better source modelling. Nevertheless, image-based subtraction of continuum residuals will likely be needed also for full ASKAP, e.g. to remove artefacts from time-variable radio sources.

Here our image-based continuum subtraction is done by fitting a first order polynomial to the spectrum of each individual pixel. We use a dynamic threshold algorithm based on the flux variation within each spectrum to exclude H I emission and other bright line signals. It is important that all the continuum emission is subtracted, without over-subtracting sources with strong H I emission that would affect the spectral shape or flux of the source. For IC 5201 we have tested that the bandpass after continuum subtraction is indeed flat and does not have any negative artefacts or bowls around the bright emission.

### 5.1.6  *Mosaicking*

The spectral cubes created for each ASKAP beam are mosaicked together to create a single large cube. The coaddition is weighted in proportion to the number of visibilities used to make that spectral cube. The mosaicking process includes primary beam correction, where we assume that the beam shape is a circular 2D Gaussian of FWHM = 1° at 1.4 GHz.

Due to the interleave pattern of the two (main) footprints, there should be approximately uniform sensitivity across the overlapping regions of the mosaic. IC 5201 is intentionally located in the centre of the mosaic, where there is the most overlap and thus the best sensitivity. A typical RMS noise for our mosaic is 2 mJy beam$^{-1}$ per 4 km s$^{-1}$ channel. In the most sensitive region of our image, our RMS noise is 1.7 mJy beam$^{-1}$ per 4 km s$^{-1}$ channel, which is the expected sensitivity of the full WALLABY survey. The pixel size of our final mosaic is 5 $''$ and the resultant synthesized beam 23.7 $'' \times 18.2$ $''$. This results in a $3\sigma$ H I column density of $1.4 \times 10^{20}$ cm$^{-2}$ over 25 km s$^{-1}$. The mosaicked image cube of IC 5201 is publicly available and can be downloaded from the CSIRO ASKAP Science Data Archive (https://doi.org/10.25919/5d1ed28479f82).

### 5.2  DECam images and photometry

We use the DES image cutout service and database access to query the DR1 catalogue. For the galaxies detected in our ASKAP image, we extract a 12′ × 12′ cutout from the DES co-added images, providing images in all the *g*, *r*, *i*, *z* and *Y* bands. We create 3 colour images from the *g* (blue), *r* (green) and *i* (red) –band images. For IC 5201, multiple cutouts are required, that are first mosaicked together prior to making the 3-colour image.

Additionally, we query the database using the known coordinates of the galaxies to identify and extract the photometry in the same bands for each galaxy from the DES DR1 catalogue. We refer the reader to Morganson et al. (2018) for the imaging processing pipeline and Abbott et al. (2018) for the details on the photometry in the DES DR1 catalogue.

### 5.3  *GALEX* images and photometry

We obtain the most recent *GALEX* data release (GR6/7) NUV and FUV images from the Mikulski Archive for Space Telescopes (MAST). The photometry for our sample does not come from the archive catalogues which result from the automated pipeline but have been individually remeasured in the same manner as for Wyder et al. (2005); Meurer et al. (2009); Wong et al. (2016); Wang et al. (2017); Audcent-Ross et al. (2018). Due to the large 1.5-degree *GALEX* field-of-view, the automated background subtraction is not accurate enough for the purposes of obtaining accurate photometry in individual galaxies. The main benefit of our re-measured photometry is the background estimate of the region specific to each galaxy.

From the extracted image, foreground stars and background galaxies were masked and a curve of growth was measured, where the point that it flattened off was used to determine an appropriate aperture to measure the photometry. The background is determined from an outer annulus beyond the galaxy's aperture. Finally, NUV and FUV photometry was measured for the galaxies.

The final FUV luminosities include the correction for Galactic reddening and internal dust correction as described in Wong et al. (2016). The uncertainties estimated for the sample include systematic uncertainties in the measurement but not that of the internal dust correction.

## 6  ANALYSIS

To find H I sources in our mosaicked WALLABY data cube we use our Source Finding Application (SoFiA: Serra et al. 2015). Within SoFiA, we first use the *filterArtefacts* algorithm that identifies and flags sightlines affected by continuum artefacts, that was applied prior to the default *Smooth + Clip* algorithm. To obtain a reliable and highly complete source catalogue, we chose a $4\sigma$ detection threshold using the local noise estimate, with the reliability filtering (as described by Serra et al. 2012) switched on. All sources with a reliability $\geq 85\%$ were found to be real and this threshold did not include any spurious detections in the output catalogue. The flux of each source was complete to the level where (further) mask dilation was not required. Any missing flux that would have been included from applying the mask dilation is included in our uncertainties.

The H I intensity map, velocity field, velocity dispersion (moment 0, 1 and 2 respectively) and integrated spectra are produced from SoFiA. The position, $v_{sys}$, $w_{20}$ and H I flux density are determined from the SoFiA catalogue. The H I mass is calculated in the standard way, following

$$\left( \frac{M_{HI}}{M_\odot} \right) \approx 2.36 \times 10^5 \left( \frac{D^2}{Mpc} \right) \int \left[ \frac{S(v)}{Jy} \right] \left( \frac{dv}{km \ s^{-1}} \right) \quad (1)$$

where D is the distance to the galaxy in Mpc, and the flux density $S(v)$ in Jy is integrated over the full velocity range where there is H I emission, with $dv$ as the velocity resolution in km s$^{-1}$. The uncertainty of $v_{sys}$ is calculated in the same way as Koribalski et al. (2004), that takes into account the shape of the H I profile. The uncertainty in $w_{20}$ is taken from the SoFiA catalogue and the uncertainty of the H I flux density is measured from the combination of the uncertainty of each point in the integrated spectra, the additional flux from the SoFiA mask dilation and the average ASKAP-SUMSS flux ratio uncertainty. The uncertainty of the flux density at each channel in the spectrum is determined by the square root of the number of independent pixels that contribute to that channel. These are propagated to the H I mass uncertainty and subsequent quantities that use the H I mass.

In order to further parameterise the sources detected by SoFiA we cut out smaller cubes around their central location that comfortably include the detections. These cubelets are then fitted with the Fully Automated TiRiFiC (FAT; Kamphuis et al. 2015) which operates directly on the 3D cubelets. FAT is an automated wrapper around the Tilted Ring Fitting Code (TiRiFiC; Józsa et al. 2007) and it first obtains the initial estimates for the fit from SoFiA. It then fits a symmetric flat, i.e. no radial variations of position angle (PA) and inclination, tilted ring model (TRM) to the data and subsequently fits a final model where the rotation curve is kept circularly symmetric but the inclination and PA are independently fitted for the approaching and receding side allowing for radial variations in order to account for any possible warping of the disc. For further details on the fitting process of FAT we refer to Kamphuis et al. (2015).

As IC 5201 is part of the SINGG and SUNGG sample (Meurer et al. 2006; Wong et al. 2016), there exist stellar mass and SFR measurements that we use in this work. The stellar mass was estimated from the SINGG $R$-band observations and the SFR was measured from the SUNGG FUV luminosity. Further details on the comparison between the estimations of SFR using the FUV versus H$\alpha$ can be found in Audcent-Ross et al. (2018) and Meurer et al. (2009). In general, galaxies with lower star formation rates are better traced by the FUV than by H$\alpha$, because the former corresponds to less massive and larger number of stars, and hence better samples the IMF than the latter (Calzetti 2013).

For all other galaxies detected in the ASKAP cube, we use the $g$ and $i$ photometry extracted from the DES DR1 catalogue to estimate stellar masses. To estimate stellar masses, we use the empirical relation from Taylor et al. (2011):

$$\log\left(\frac{M_\star}{M_\odot}\right) = 1.15 + 0.7(g-i) - 0.4M_i \qquad (2)$$

where $g$ and $i$ are the apparent magnitude in each filter and $M_i$ is the absolute $i$-band magnitude. This relation was derived from SED modelling of galaxies in $g$ and $i$ filters for GAMA galaxies, that are the same $g$ and $i$ filters on the DECam. The relation is accurate within 0.1 dex (Taylor et al. 2011), which we use as our uncertainty.

The SFR for all the galaxies in our sample with FUV photometry are calculated assuming a Kroupa initial mass function (IMF; Wong et al. 2016) using

$$\left(\frac{SFR_{FUV}}{M_\odot \ yr^{-1}}\right) = \frac{1}{1.37 \times 10^{32}}\left(\frac{l_{FUV}}{W \text{Å}^{-1}}\right) \qquad (3)$$

where $l_{FUV}$ is the dust-corrected (section 5.3) FUV luminosity. For any galaxy without FUV photometry, we estimate the SFR using the NUV luminosity which is only corrected for Galactic reddening using the method described by (Schiminovich et al. 2007). As both the FUV and NUV emission are probing slightly different star formation timescales and stellar populations, a comparison of the NUV-derived SFR to that obtained from the FUV (for the sample of galaxies for which both FUV- and NUV-luminosities are available) reveal that on average, the NUV-derived SFR appears to be a factor of 2.7 greater than that from the FUV. We include this factor in our uncertainties and remind the reader to be aware of this difference when comparing SFR between the galaxies in this sample.

For all the galaxies, we measure the H I fraction, defined as the H I-to-stellar-mass ratio, the specific star formation rate (SSFR), defined as the SFR-to-stellar mass ratio and the star formation efficiency (SFE), defined as the SFR-to-H I mass ratio and is the efficiency that a galaxy will use its H I to form stars at the current SFR.

## 7   RESULTS

We report a total of 9 extragalactic H I detections in the mosaicked ASKAP cube (Fig. 3), where four of these are new H I detections. Eight of the H I detections have clear optical counterparts and the remaining H I detection is a tidal feature. We detect the barred spiral galaxy IC 5201 and the three, gas-rich dwarf galaxies, AM 2220−460, ESO 289-G020, 6dF J2218489−461303 (hereafter 6dF J22) at the systemic velocity of IC 5201. We also detect GALEXASC J222217.98−464204.7 (hereafter GALEXASC J22) and ESO 289-G05 that have the same systemic velocity as the NGC 7232/3 group. In the central region of the NGC 7232/3 group, we detect, NGC 7232B, NGC 7233 and a tidal feature, consistent with C6 in Lee-Waddell et al. (2019).

These are the first H I detections for AM 2220−460, ESO 289-G020, 6dF J22 and GALEXASC J22, providing the first systemic velocity, and thus redshift (cz) measurements for AM 2220−460 and GALEXASC J22. We present the position, systemic velocity, width of the spectral profile at 20% of the peak flux ($w_{20}$), H I flux density and H I mass in Table 3, as well as the optical photometry, UV photometry, stellar mass and SFR in Table 4 for all our detections except those in the NGC 7232/3 triplet. We do not include calculations of the NGC 7232/3 triplet as it is an extended and diffuse system, located at the edge of our 4-beam image. The sensitivity is the lowest at the edges of the images and we refer the reader Lee-Waddell et al. (2019) who do a detailed study of the triplet.

### 7.1   IC 5201 and its satellites

The H I images of IC 5201 in this work are the highest resolution images to date in both the spatial and spectral domain.

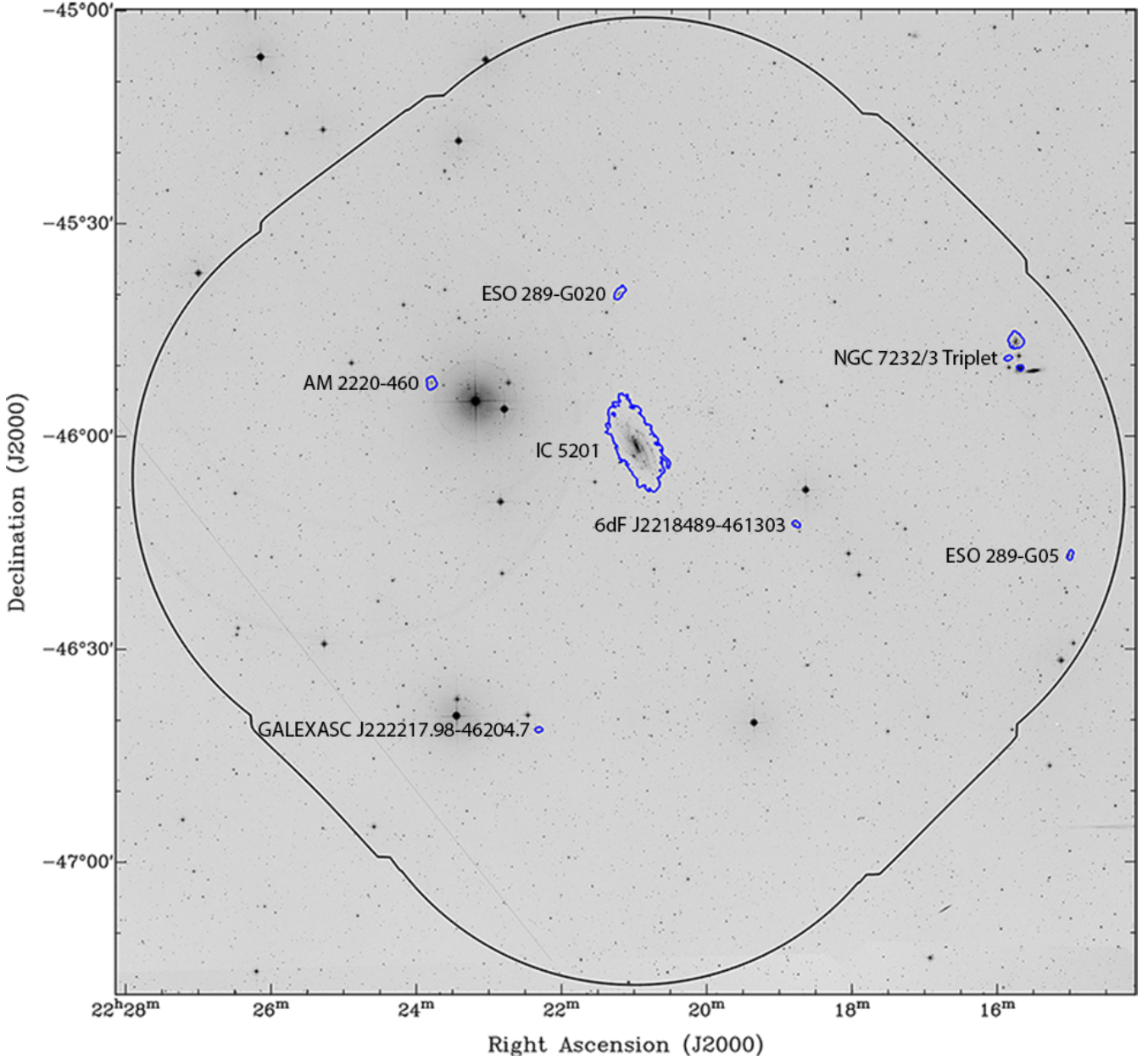

**Figure 3.** The ASKAP H I sources (blue contours), as detected using `SoFiA`, overlaid onto an optical DSS2 *B*-band image. The $3\sigma$ column density level (blue contours) is $\simeq 1.4 \times 10^{20}$ cm$^{-2}$. The black outline shows the border of the mosaicked cube. In total, we detect 9 ASKAP H I sources: 8 galaxies and a tidal features in the NGC 7232 triplet.

Using these in conjunction with the detection and images of its satellites, the pertinent aim of this work is to determine what physical conditions have resulted in the observed H I, stellar and star formation distributions for IC 5201 and its satellites.

We present the channel maps of IC 5201 in Fig. 4, that show the rotation of the H I disc. The channel maps reveal that there is H I emission that deviates from the expected emission for a flat, rotating disc, indicating a warped H I disc. To visualise the warps, we show an interactive, 3D rendered representation of the H I disc of IC 5201 (Fig. 5), created using Slicer Astro (Punzo et al. 2017). This figure can be controlled using Adobe Acrobat and it is clear in this representation that the H I disc of IC 5201 is warped.

We quantify the kinematics of IC 5201 and present the rotation curve and the variation of disc inclination and position angle as a function of radius (Fig. 6). The rotation curve flattens at $v_{\rm rot} = 98$ km s$^{-1}$ for large radii. At a radius of 14 kpc, $\sim 2$ kpc before the edge of the stellar disc, both the approaching and receding side of the disc start becoming warped. The warp increases and becomes more prominent at larger radii, in the outer disc, where the receding side is more warped than the approaching side. The largest difference between the two sides is 20° in the position angle at a radius of 25 kpc.

**Table 3.** Properties of the H I detections (excluding the NGC 7232/3 triplet). The position, $v_{sys}$, $w_{20}$ and H I flux density are taken directly from the `SoFiA` catalogue and the H I mass is calculated in the standard way using the flux density and distance. Uncertainties in the systemic velocity are calculated the same as Koribalski et al. (2004) and uncertainties in $w_{20}$ is from the `SoFiA` catalogue. The H I flux density uncertainty is a combination of the uncertainty on the flux along the line of site in each point of the spectrum, the uncaptured flux (from mask dilation) and the ASKAP-SUMSS flux ratio, where this uncertainty is propagated to the uncertainty of the H I mass.

| ASKAP ID | Source | RA, Dec (J2000) | $v_{sys}$ (km s$^{-1}$) | $w_{20}$ (km s$^{-1}$) | Flux Density (Jy km s$^{-1}$) | $M_{HI}$ ($10^8$ M$_\odot$) |
|---|---|---|---|---|---|---|
| WALLABY J222057−460201 | IC 5201 | 22:20:57.9, −46:02:01.2 | 914 ± 2 | 205 ± 2 | 177 ± 1 | 73.1 ± 0.3 |
| WALLABY J222343−456306 | AM 2220−460 | 22:23:43.0, −45:53:06.2 | 877 ± 2 | 70 ± 4 | 1.9 ± 0.2 | 0.7 ± 0.1 |
| WALLABY J222111−454033 | ESO 289-G020 | 22:21:11.5, −45:40:33.5 | 917 ± 2 | 95 ± 5 | 2.5 ± 0.3 | 1.0 ± 0.1 |
| WALLABY J221848−461305 | 6dF J22 | 22:18:48.7, −46:13:05.3 | 996 ± 2 | 40 ± 2 | 0.6 ± 0.1 | 0.3 ± 0.1 |
| WALLABY J222217−464206 | GALEXASC J22 | 22:22:17.8, −46:42:06.2 | 1882 ± 5 | 66 ± 6 | 0.6 ± 0.1 | 1.1 ± 0.2 |
| WALLABY J221506−461658 | ESO 289-G05 | 22:15:06.3, −46:16:58.2 | 1935 ± 4 | 136 ± 15 | 1.4 ± 0.3 | 2.6 ± 0.5 |

**Table 4.** The DECam optical photmetry, *GALEX* UV photometry, stellar mass and SFR that were derived from the photometry. For IC 5201, the stellar mass and SFR were obtained from SINGG and SUNG (Meurer et al. 2006; Wong et al. 2016) and there was no available FUV photometry for ESO 289-G05.

| Source | $g$ (mag) | $i$ (mag) | NUV (mag) | FUV (mag) | $\log(M_\star)$ (M$_\odot$) | SFR (M$_\odot$ yr$^{-1}$) |
|---|---|---|---|---|---|---|
| IC 5201 | n/a | n/a | 12.95 | 13.17 | 9.8 | 63.0 |
| AM 2220−460 | 16.25 | 15.73 | 17.56 | 18.12 | 7.4 | 0.6 |
| ESO 289-G020 | 15.51 | 15.02 | 17.31 | 17.56 | 7.3 | 0.3 |
| 6dF J22 | 15.80 | 15.42 | 17.23 | 17.43 | 7.6 | 0.4 |
| GALEXASC J22 | 16.87 | 16.22 | 18.98 | 19.41 | 7.9 | 0.2 |
| ESO 289-G05 | 14.67 | 14.14 | 16.51 | n/a | 8.8 | 0.7 |

In Fig. 7 we show the integrated H I spectrum for IC 5201, AM 2220−460, ESO 289-G020 and 6dF J22, where $v_{sys}$ of all three dwarf galaxies are within 80 km s$^{-1}$ of $v_{sys}$ for IC 5201. The projected distances between IC 5201 and AM 2220−460, ESO 289-G020, 6dF J22 are 115, 83 and 96 ± 5 kpc respectively, assuming that all galaxies are at a distance of 13.2 Mpc. Both the stellar and H I mass of IC 5201 are ∼ 2 orders of magnitude larger than those of the three dwarf galaxies. The apparent close proximity and relative masses of the dwarf galaxies signify that IC 5201 is the centre of the system and the rest are satellites. These are the first reliable H I detections of the three satellites and we add AM 2220−460 as a new satellite, given there was no previous distance measurement to it. Our measurements of $v_{sys}$ for ESO 289-G020 and 6dF J22 agree with and improve the precision compared to the spectroscopic redshifts in the literature.

In Fig. 8, we show the three-colour DECam optical image overlaid with ASKAP H I contours, the H I column density map derived directly from the H I intensity (moment 0) map, the H I velocity field (moment 1 map) and H I velocity dispersion (moment 2 map) for all galaxies in this system.

IC 5201 has a H I disk (∼ 25 kpc) that extends beyond the stellar and UV disk (∼ 14 kpc) along the projected major axis in the North-East and South-West directions. The densest (i.e. $N_{HI} > 2 \times 10^{21}$ cm$^{-2}$) H I in IC 5201 directly coincides with the dense blue star clumps in the optical image as well as the knots of UV emission. Therefore, the site of current star formation still has a reservoir of H I that could fuel and sustain further star formation. The velocity field shows the H I rotation in IC 5201, and we further assessed the H I emission to quantify the warped morphology and kinematics (Fig. 5 and 6).

The optical morphology of AM 2220−460 is irregular and clumpy, with the H I emission being similar. The velocity field shows a bent iso-velocity contour along the minor axis (Fig. 8), though it is not resolved enough to quantify the kinematics. This is the most likely galaxy to show evidence of a previous interaction with another galaxy.

ESO 289-G020 is an edge-on galaxy with an extended H I disc. The velocity field (Fig. 9) appears undisturbed and regularly rotating, which is also apparent given the relatively symmetric double horn profile (Fig. 7). There is no evidence that there have been any significant interactions with ESO 289-G020 in this system.

6dF J22 appears to be triaxial galaxy in the optical image (Fig. 8). It has a similar stellar mass to the other satellites, however its H I mass is a factor of ∼ 2-3 lower. The H I profile of 6dF J22 is very narrow, with $w_{20} = 40 \pm 2$ km s$^{-1}$ (Fig. 7, Table 3), and the velocity field does not show any clear signs of rotation. While it is different from the other edge-on galaxy, the stellar mass, H I mass and $w_{20}$ of 6dF J22 are consistent with the physical properties of small, triaxial galaxies in the literature (Hunter et al. 2012; Oh et al. 2015).

We use Fig. 9 to illustrate the H I and star formation properties of each galaxy in Table 3, to search for signatures of unusual star formation (i.e. starburst or quenching). As robust scaling relations do not currently exist down to the mass of the satellites, to first order, we are searching for an ∼ order of magnitude difference between the galaxies in the H I and SFR ratios.

For IC 5201 and its satellites, the scatter in the H I fraction, SSFR and SFE is 0.3, 0.5 and 0.6 dex respectively, where no galaxy stands out in any of the metrics. The irregular satellite AM 2220−460 has the highest H I fraction and SSFR in the system, though it is not statistically significant above the scatter. All galaxies in this system have a similar ratio between the H I gas, star formation rate and stellar mass. The mean H I velocity dispersion ($\sigma_{HI}$) of these galaxies are consistent with the typical values observed in nearby spiral galaxies (e.g. van der Kruit & Shostak 1982; van Zee & Bryant 1999).

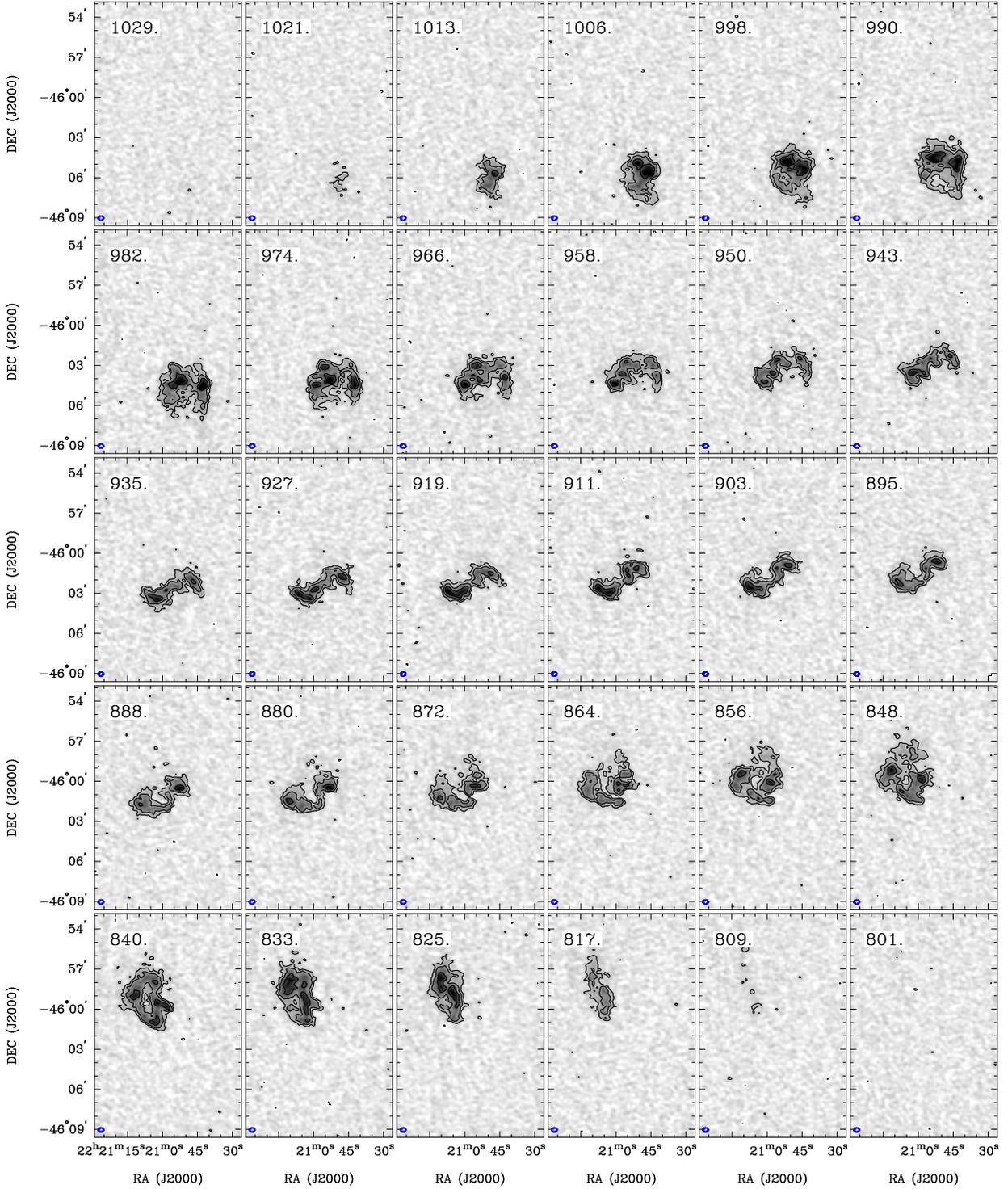

**Figure 4.** ASKAP H I channel maps of the barred spiral galaxy IC 5201. For display purposes we average the channels to show 8 km s$^{-1}$ wide channels. The contour levels are -5, 5, 10, 20 mJy beam$^{-1}$. The heliocentric velocity is shown in the top left corner while the synthesized beam (blue) that is $23.75'' \times 18.15''$ is shown in the bottom left corner. The channels (e.g. $v = 927$ km s$^{-1}$) with emission that deviates from the expected H I emission for a flat, rotating disc, shows evidence of a warped H I disc.

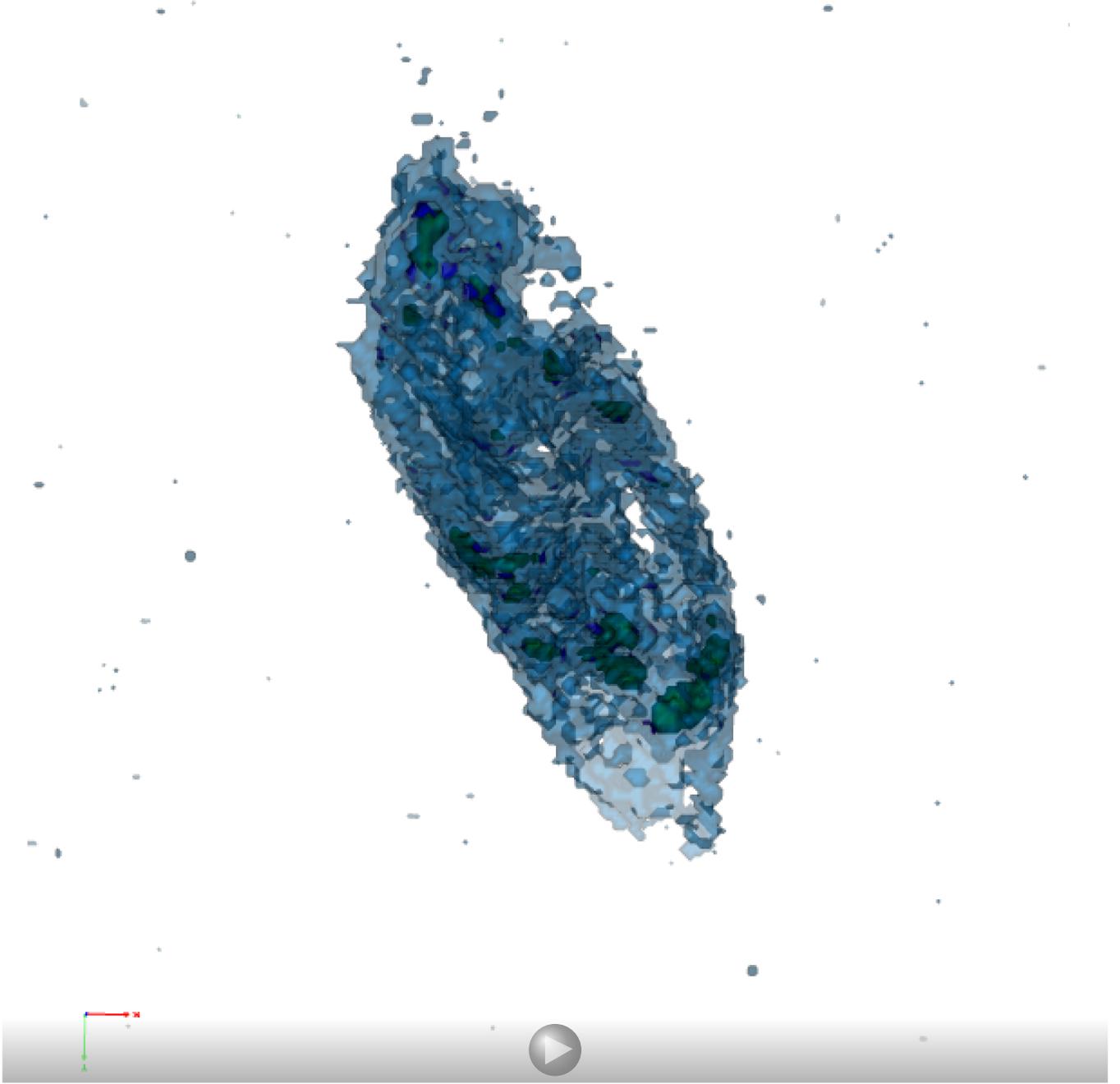

**Figure 5.** An interactive 3D representation of the H I disk of IC 5201. The figure was created with Astro Slicer (Punzo et al. 2017) and can be controlled when opened in Adobe Acrobat. The x, y, and z (red, green and blue) axes correspond to $\alpha$, $\delta$, velocity. There are three (light blue, dark blue and green) surfaces that represent three different H I densities. They are rendered at $N_{HI} \sim 10$, 20 and 30 $\times 10^{20}$ cm$^{-2}$, such that the colours are consistent with the H I column densities in Fig. 8. These three surfaces highlight the warped H I disk, the H I densities that correlate with the stellar disk as well as the H I densities coinciding with the knots of star formation.

### 7.2 The NGC 7232/3 triplet, ESO 289-G05 and GALEXASC J22

We detect two galaxies and a tidal feature in the NGC 7232/3 triplet ($v_{sys} = 1915$ km s$^{-1}$), however we refrain from quantitative measurements of these detections as the system is extended, diffuse and located in the least sensitive area of our image. We remind the reader to see Lee-Waddell et al. (2019) for ASKAP images and analysis of NGC 7232/3.

We detect an additional two galaxies at the same velocity of the NGC 7232/3 triplet. Assuming these galaxies are at the same distance as the triplet (27.3 Mpc), ESO 289-G05 and GALEX J22 are at projected distances of 208 and 677 ± 15 kpc respectively, from the group centre. The H I spectra of these two galaxies are shown in Fig. 10, and the three-colour

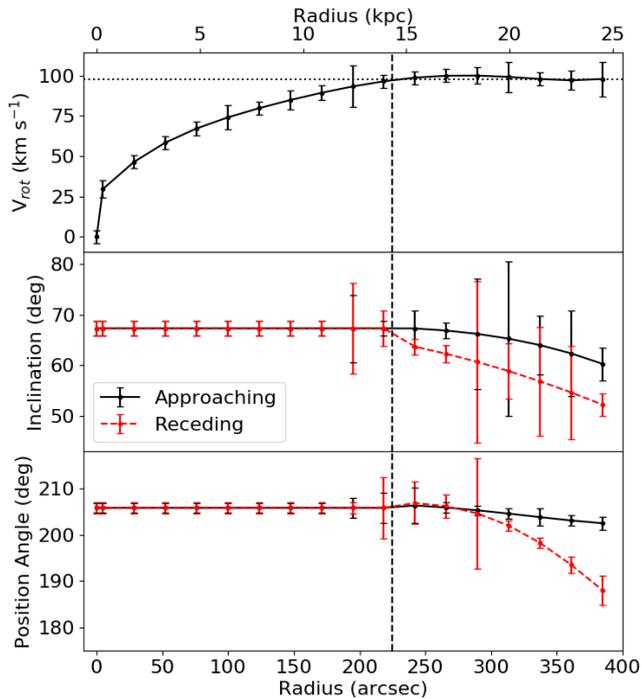

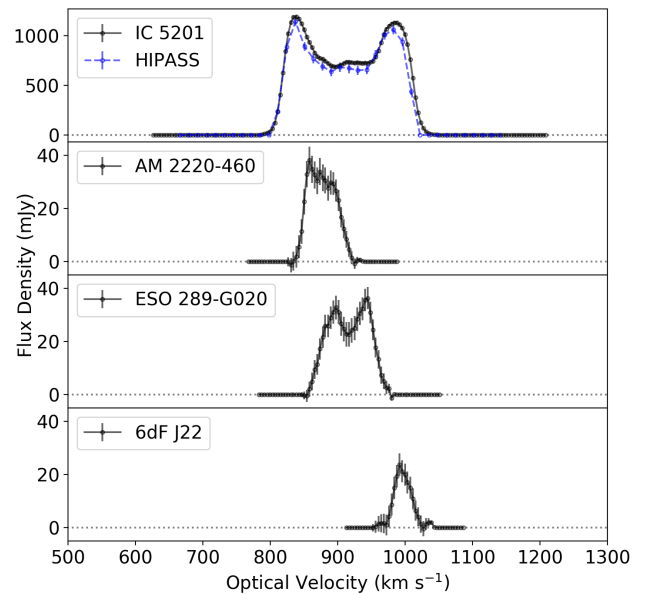

**Figure 6.** The results of the kinematic fits for IC 5201 using FAT (Kamphuis et al. 2015), where the dashed vertical line shows the limit (∼ 14 kpc) of the bright stellar disc. The top panel shows the rotation curve and the dotted line is a constant velocity of v = 98 km s$^{-1}$, the value where the profile flattens. The middle panel shows how the inclination along the major axis varies as a function of radius, where the approaching side is represented by the black solid line and the receding side is the red dashed line. The bottom panel shows the inclination and PA as a function of radius, where the approaching and receding side are the same colour and style as above. The inclination and PA starts to deviate between approaching and receding sides at a radius of 14 kpc, becoming more apparent at larger radii. Therefore, the disc becomes more warped in the outer regions, which is most easily seen in the PA, with the greatest difference of ≈ 20° at 25 kpc.

**Figure 7.** The ASKAP HI spectra of IC 5201 and its satellites, showing the flux density as a function of optical velocity (cz), where each spectra has been smoothed using a boxcar order 3 and the uncertainties are derived from the number of channels contributing flux in each point. In the top panel we overlay the HIPASS spectrum (blue dashed) of IC 5201 as a comparison, extracted using SoFiA and uncertainties derived in the same way as the ASKAP spectrum. There is excellent agreement with the peak and integrated flux between the ASKAP and HIPASS spectra.

DECam optical image overlaid with ASKAP HI contours, the HI column density map and HI velocity field are shown in Fig. 11.

Both ESO 289-G05 and GALEXASC J22 are new HI detections. ESO 280-G05 is isolated (compared to the triplet) and is also detected in Lee-Waddell et al. (2019), therefore providing an ideal test for independently created ASKAP HI images, that we discuss in the next section. As this is the first HI detection of GALEXASC J22, it provides the first distance measurement from its systemic velocity. At a projected distance of 677 kpc from the triplet, GALEX-ASC J22 is possibly an infalling member or a new group member.

The stellar emission of ESO 289-G05 is closely traced by the HI emission, that shows rotation. The HI emission of GALEXASC J22 extendes beyond the stellar disk, although it is almost unresolved (Fig. 11). The HI fraction, SSFR and SFE of these galaxies are within the scatter and consistent with all the other detected galaxies, showing no obvious evidence of a recent starburst or quenching of star formation (Fig. 9).

# 8 DISCUSSION

The ASKAP measurements of HI flux, peak flux, $v_{sys}$ and $w_{20}$ of IC 5201 (Table 3), from the spectrum (Fig. 7), all agree with those measured from the HIPASS Bright Galaxy Catalogue (Koribalski et al. 2004). This shows that ASKAP is recovering the (correct full amount of) flux and that the ASKAPsoft data reduction pipeline is working well for IC 5201, a bright, extended HI source. For early science, some processing is required outside the pipeline, but that will become redundant as more antennas are added to the array along with new ASKAP observing techniques and improvements to ASKAPsoft.

At least two-thirds of disc galaxies in the local Universe contain a bar (e.g. Cheung et al. 2013; Buta et al. 2015). The bar can alter the angular momentum of the galaxy and funnel stars and gas towards the centre, increasing the SFR (e.g. Muñoz-Mateos et al. 2013). There is no clear correlation between the central bar and the HI intensity. The bar is visible in the HI velocity field as the blue and red shifted (810 and 1100 km s$^{-1}$ respectively) bar in the centre along the major axis (Fig. 8). The highest $\sigma_{HI}$ occurs at the same location. This could suggest that H I is being funneled along the bar, however, this does show clear causation between the bar and $\sigma_{HI}$, as the highest $\sigma_{HI}$ typically occurs in the centre of galaxies (Fig. 8 and 11).

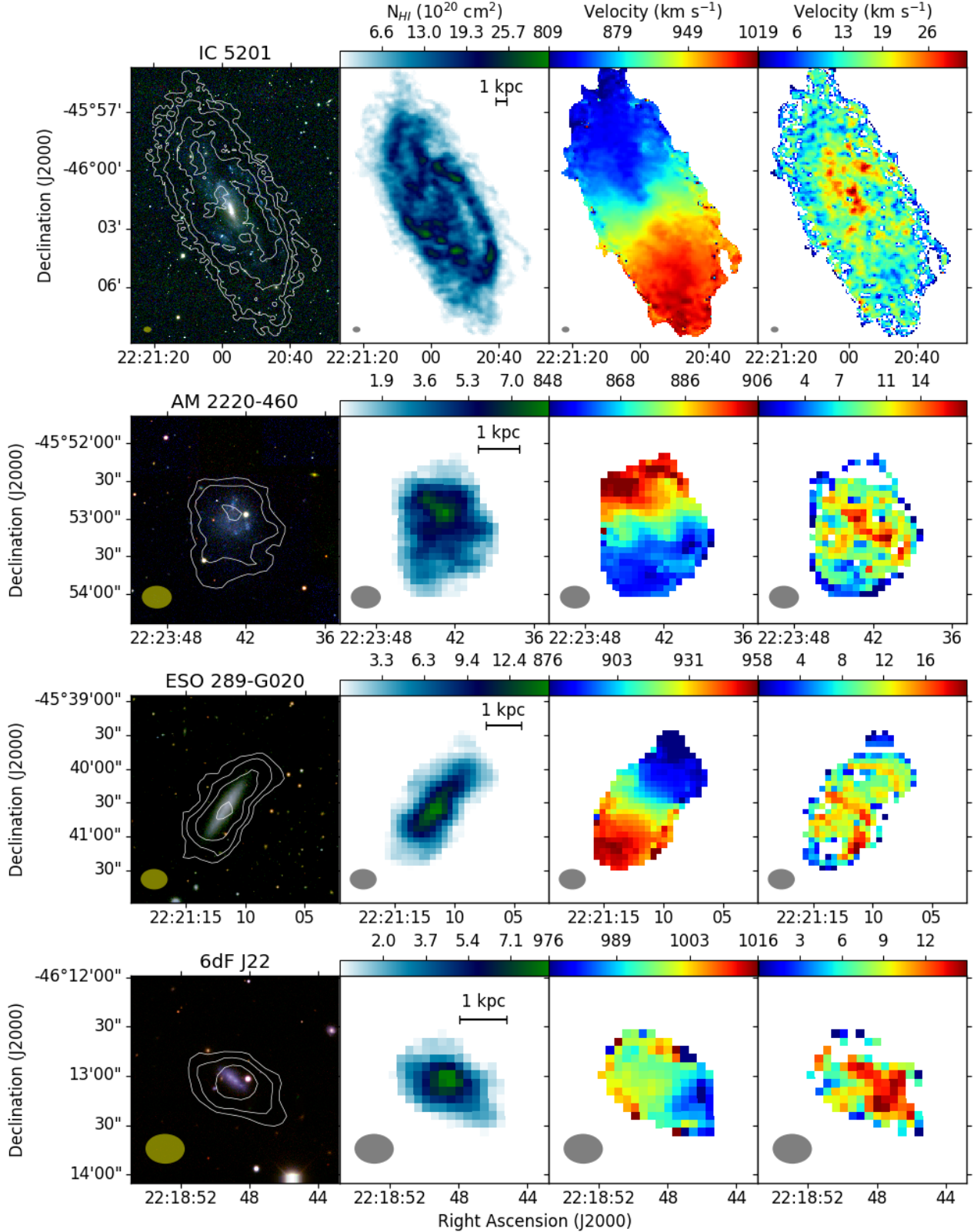

**Figure 8.** The DECam optical and ASKAP H I images for IC 5201 and its satellites. The left panel shows the 3-colour composite, created from the $g$-, $r$- and $i$-band filters with (white) H I contours overlayed. For IC 5201, the contour levels are $3\sigma \times 3^n$, while the rest of the galaxies are $3\sigma \times 2^n$, where $\sigma$ is the RMS noise level in each image and $n = 1, 2, 3, 4$. The second panel is the H I intensity (moment 0) map that has been directly converted into H I column density, the third panel shows the H I velocity field (moment 1 map) and the fourth panel shows the H I velocity dispersion (moment 2 map). In each image, the synthesized ASKAP beam is shown on the bottom left, where North is up and East is to the left. The H I emission extends beyond the stellar component for all the galaxies. In the overlap, the H I typically traces the optical morphology where the H I and stellar over-densities directly coincide. IC 5201, AM 2220−460 and ESO 289-G020 all show clear rotation of the H I, though the H I in 6dF J22 does not. See the text for further discussion.

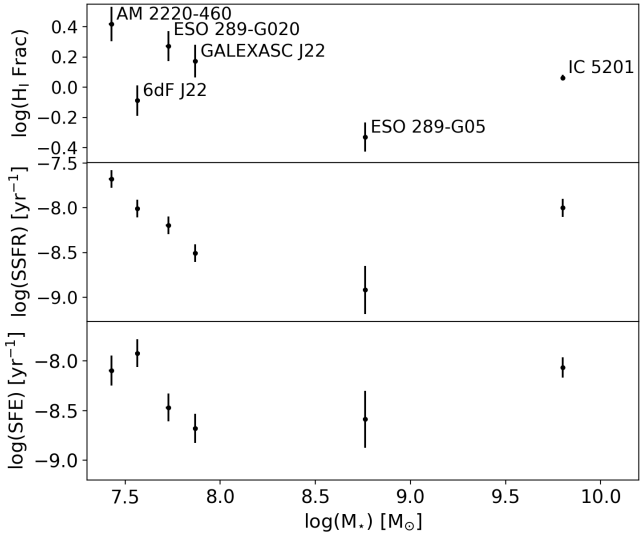

**Figure 9.** The H I-to-stellar mass ratio (top), star formation rate-to-stellar mass ratio (middle) and star formation rate-to-H I mass ratio (bottom) as a function of stellar mass for 6 galaxies in the ASKAP cube. The H I masses were measured from the ASKAP spectra, SFRs from *GALEX* photometry and stellar masses from DECam photometry. These metrics are used to search for strong interactions that result in galaxies with unusual properties i.e. galaxies undergoing a starburst or quenching. Despite the range of galaxy (H I and stellar) mass, morphology and environment, all of these galaxies display similar H I gas, SFR and stellar mass ratios.

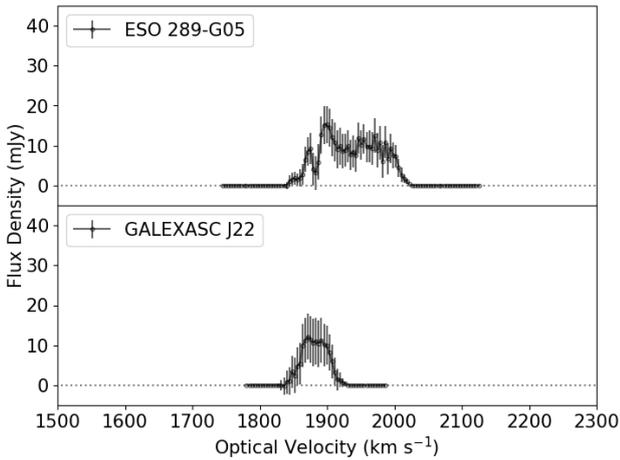

**Figure 10.** Same as Fig. 7, showing the ASKAP H I spectrum for ESO 289-G05 and GALEXASC J22. Both galaxies are clearly detected, although the noise of these spectra are a factor of $\sim 2$ higher than the galaxies detected near the centre of the image. These galaxies are likely to be new or infalling members of the NGC 7232/3 group.

There are some signatures that IC 5201 may have been disturbed in the past. The H I disc is warped and the optical morphology shows asymmetries between the spiral arms (Fig. 1). Relative to the average SFE of nearby star-forming galaxies from the SUNGG sample (Wong et al. 2016), IC 5201 has an SFE that is a factor of 2.5 lower.

The mean H I velocity dispersion of IC 5201 is $\sigma_{HI} = 13.6$ km s$^{-1}$, as measured directly from the moment 2 map and FAT. Considering that $\sigma_{HI} > 11$ km s$^{-1}$, it is greater than the average $\sigma_{HI}$ for nearby (SINGG/SUNGG and THINGS) galaxies (Leroy et al. 2008; Zheng et al. 2013; Wong et al. 2016). This suggests that the low SFE of IC 5201 is likely to be due to the galaxy's past interaction history, opposed to galaxies being highly inefficient in converting gas to stars because of secular properties such as very stable disks (Wong et al. 2016) or very high angular momentum (Lutz et al. 2017, 2018). This is not definitive evidence that IC 5201 has a non-secular interaction history, as we use the same assumptions and disk stability model used for the SINGG/SUNGG and THINGS galaxies (Leroy et al. 2008; Zheng et al. 2013; Wong et al. 2016), that falsely inflate $\sigma_{HI}$ from beam smearing and projection effects. However, we present and discuss further evidence for this scenario later in this section.

All dwarf satellite galaxies within 250 kpc of the Milky Way do not have any detected H I down to $\sim 10^5$ M$_\odot$ (Spekkens et al. 2014). The exceptions are the Small and Large Magellanic Clouds, that have H I masses of 3.8 and 4.8 $\times 10^8$ M$_\odot$, respectively (Stanimirovic et al. 1999; Staveley-Smith et al. 2003). In contrast, all detected dwarf satellites of IC 5201 are within a projected distance of 120 kpc, have H I masses between $0.3 - 1.0 \times 10^8$ M$_\odot$ and a $V$ band luminosity (and hence, comparable stellar mass) to the Sagittarius dSph satellite of the Milky Way (Spekkens et al. 2014). This implies that there is a critical mass and/or radius for satellites to retain their cool gas. The driving factors would be the mass of the central galaxy and the mass ratio between the satellites. As IC 5201 is approximately 4 times less massive than the Milky Way, dwarf H I rich satellites can retain their cool gas in closer proximity to IC 5201.

We search for strong signatures of interactions between IC 5201 and its satellites. As a first-order test, we test for any unusual H I fractions, SSFRs and SFEs (Fig. 9), that would highlight strong processes directly affecting the H I content and star formation. In the IC 5201 group, no galaxy stands out with an unusual H I fraction, SSFR or SFE as a result of strong interactions. In addition to this, there are no tidal streams or H I clouds between the galaxies down to a column density of $N_{HI} = 1.4 \times 10^{20}$ cm$^{-2}$ (Fig. 3). H I tidal streams and clouds are detectable at this column density (e.g. the M81 triplet; de Blok et al. 2018) and have been detected using the same ASKAP observations (the NGC 7232/3 triplet; Lee-Waddell et al. 2019), though the lack of intergalactic H I implies no strong tidal interactions between IC 5201 and its satellites, that is a different dynamical history compared to these other systems.

The minimum stellar mass ratios for the the M81 and NGC 7232/3 systems are $\approx 26$ and 2 (Karachentsev et al. 2013; Lee-Waddell et al. 2019), while the minimum stellar mass ratio for the IC 5201 group $\approx 117$. As there are 1-2 orders of magnitude difference in the mass ratios, the tidal interactions in the other systems are much stronger and we do not expect the satellites of IC 5201 to be able to tidally disturb IC 5201 and create debris with a H I column density of $10^{20}$ cm$^{-2}$. It is possible that there are H I streams and clouds below this column density (i.e. $< 10^{20}$ cm$^{-2}$), however, more sensitive observations would be required to test if they are present.

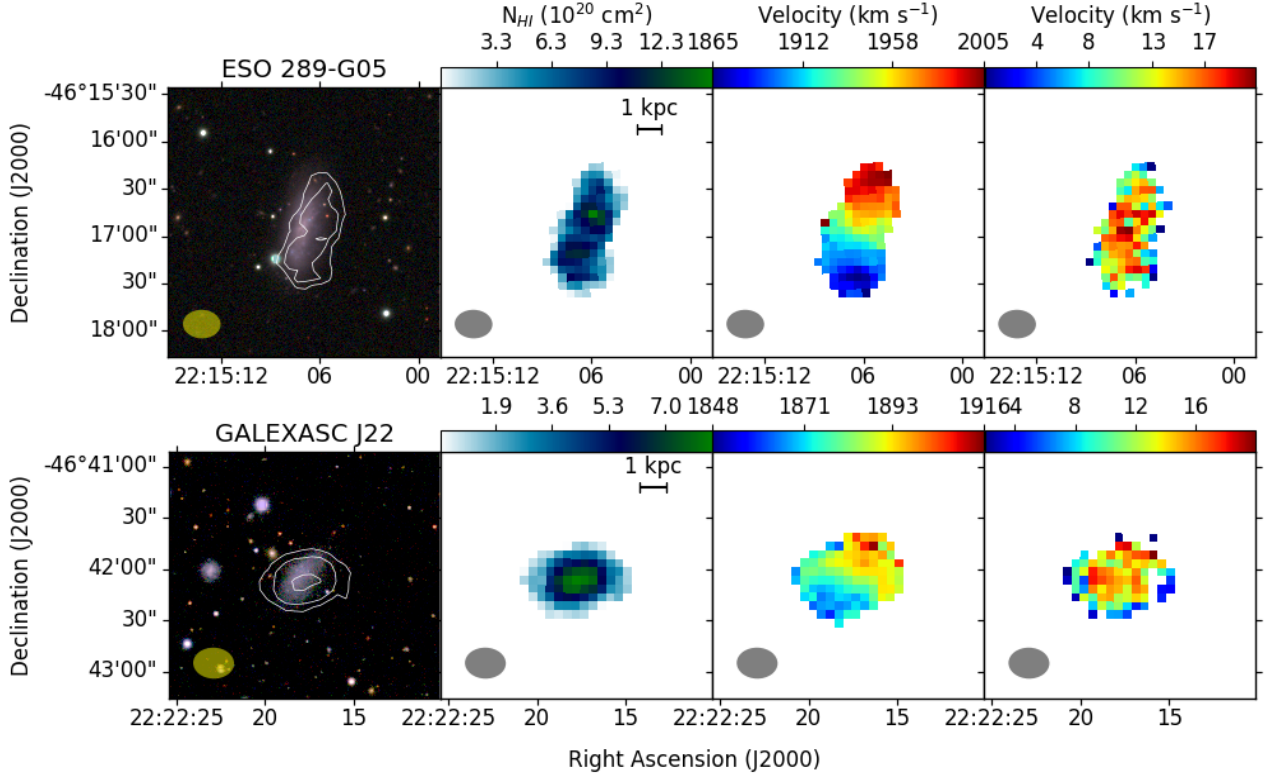

**Figure 11.** Same as Fig. 8, showing the DECam 3-colour optical and ASKAP H I images for ESO 289-G05 and GALEXASC J22. The H I (white) contours overlaid on the optical image are at the levels of $3\sigma \times 2^n$, where $\sigma$ is the RMS in each image and $n = 1, 2, 3, 4$. The H I emission closely traces the stellar emission for ESO 289-G05 and the extended H I emission for GALEXASC J22 is most likely beam smearing. There is clear rotation in the velocity field of ESO 289-G05 and it is likely that GALEXASC J22 is rotating, although the velocity gradient is less distinct. These galaxies are not well resolved with the ASKAP synthesized beam and would require higher resolution imaging to determine the H I distribution, morphology and velocity.

There is significant evidence that there has been a previous tidal interaction between IC 5201 and AM 2220−460. The asymmetric spiral arms (Fig. 1) and warped H I disc of IC 5201 (Fig. 4, 5 and 6) is likely a result of an external tidal interaction. We attempted to measure the kinematics of the satellites, however, none of them are resolved enough to do so. All three satellites are at a similar projected distance, with ESO 289-G20 at 83, 6dF J22 at 96 and AM 2220−460 at 115 ± 5 kpc. However, only AM 2220−460 has a clumpy, irregular optical and H I morphology as well as a distorted iso-velocity contour (Fig. 8). This makes AM 2220−460 the most likely candidate to have interacted with IC 5201. While it remains unclear what physical processes cause warps, large warps are usually associated with tidal interactions and small warps associated with gas accretion (Ann & Park 2006; Ann & Bae 2016). The warp of IC 5201 is significant and the irregular morphology of AM 2220−460 is consistent with tidal interactions, not gas accretion, being the cause.

For example, a possible scenario is that AM 2220−460 did a close fly-by of IC 5201, warping the stellar and H I disc of the central galaxy, while disrupting its own stellar and H I distribution into an irregular morphology. As a consistency check for this scenario, we compare the orbital period ($t_{orbit}$) of IC 5201 to the timescale that it would take AM 2220−460 to reach its current position after its close fly-by ($t_{fly-by}$). The orbital period was calculated using the de-projected radius of the spiral arm and the rotation velocity (from Fig. 6) at that radius, resulting in $t_{orbit} = 1.1$ Gyr. $t_{fly-by}$ was calculated using the rotational velocity of IC 5201 and assuming that the velocity of AM 2220−460 is the typical velocity derived from the velocity dispersion for a particle inside the IC 5201 halo. It was also assumed that both galaxies are at same distance and AM 2220−460 travels linearly, directly across the plane of the sky. We found that $t_{fly-by} = 1.6$ Gyr, the same order of magnitude as $t_{orbit}$. Is takes ~ 1 orbit (i.e. ~1.1 Gyr in IC 5201) for dynamic perturbations to settle (Holwerda et al. 2011), and even though simplified assumptions are used to estimate $t_{fly-by}$, given the time scales are the same order of magnitude, the proposed scenario is reasonable.

Warped discs are a common occurrence observed in many nearby galaxies, however, not all can be caused by tidal interactions (e.g. Koribalski et al. 2018). A warp can also be caused by accreting misaligned gas, a misaligned (and therefore precessing) DM halo that causes the inner part of the halo to move faster than the outer part, the galaxy precessing in a sufficiently dense IGM, or even by ram pressure (Haan & Braun 2014). While we can not rule out these mechanisms as the cause of the warped IC 5201 disc, there is no clear evidence to suggest that they are more likely than a tidal origin or that they would sufficiently explain the irregular properties of AM 2220−460. It is plausible that multiple mechanisms are working together, although we can not disentangle them or their relative contribution with these observations.

While we detect two galaxies and a tidal feature in the NGC 7232/3 triplet, using a lower threshold in SoFiA, we are able to detect NGC 7232, the third galaxy in the triplet as well as the H I cloud C5 (Lee-Waddell et al. 2019). However, this simultaneously introduces spurious sources into our output catalogue and demonstrates that we can not make a thorough, independent comparison to the findings of Lee-Waddell et al. (2019).

Just like the NGC 7232/3 triplet, ESO 289-G05 is detected at the edge of our image. However, ESO 289-G05 is a single, isolated galaxy that is not in a diffuse and extended system (like the triplet) and we measure its H I properties as an independent test to Lee-Waddell et al. (2019). All our measured values of the $v_{sys}$, $w_{20}$, H I flux and mass (Table 3) agree with Lee-Waddell et al. (2019). All values except the H I flux agree within $1\sigma$. The H I flux measured in this work is slightly (within $2\sigma$) lower, though this is expected as the RMS at this location is twice as large as the RMS in Lee-Waddell et al. (2019). $v_{sys}$ of ESO 289-G05 is within $\sim 80$ km s$^{-1}$ of $v_{sys}$ of the triplet. Assuming that they are located at the same distance of $\sim 26$ Mpc, ESO 289-G05 has a projected distance of 242 kpc from the NGC 7232/3 triplet, making it a likely group member or in the process of infalling.

The new H I detection of GALEXASC J22 presents a possible, new member of the NGC 7232/3 group. The systemic velocity ($v_{sys} = 1882$ km s$^{-1}$) is the same for GALEX-ASC J22 and the triplet, where GALEXCASC J22 is at a projected distance of 677 kpc from the triplet. As the images of GALEXASC J22 (Fig 11) remain unresolved, there is no evidence to suggest that GALEXASC J22 has interacted with the triplet yet.

# 9  CONCLUSIONS

We present results from the first WALLABY Early Science observations, obtained with an array of $10 - 12$ ASKAP antennas equipped with Mk II PAFs that have a field of view of 30 sq degrees. The observations targeted the NGC 7232/2 galaxy group and we focused on the nearby barred spiral galaxy IC 5201 and its surroundings, mainly to extensively test and enhance ASKAPsoft- the ASKAP data calibration and imaging pipeline. As our mosaicked image cube reached the full WALLABY depth, we were able to test and optimise a large range of gridding and imaging parameters, that will likely be used for the upcoming WALLABY pilot and full survey. Furthermore, we used the SoFiA software package to reliably find and parameterise H I sources, where we quantified the kinematics of IC 5201 using FAT.

We detected 9 extragalactic H I sources in our ASKAP data cube, that has the expected sensitivity of the full WALLABY survey. These sources are IC 5201, 3 H I rich dwarf satellites (AM 2220−460, ESO 289-G020, 6dF J2218489-461303), NGC 7232B, NGC 7233 and a tidal feature that is associated with the NGC 7232/3 triplet as well as GALEXASC J222217.98-464204.7 and ESO 289-G05 that are likely in-falling or new members of the group. 5 of these are new H I detections, where 2 provide the first velocity measurements and therefore distance estimates. Our main findings are summarised as follows:

(i) The H I images of IC 5201 are the highest resolution in the image and spectral domain, to date. The H I emission shows an extended, warped disc and coincides with the optical and UV emission. The H I flux, $v_{sys}$ and $w_{20}$ measured by ASKAP agree with the literature values.

(ii) The kinematics of the IC 5201 H I disc show that the rotational velocity flattens at $v_{rot} = 98$ km s$^{-1}$ and the warp begins at 14 kpc along the major axis, and continues to get stronger with increasing radii.

(iii) We detect 3 gas-rich dwarf satellites of IC 5201, that consist of an edge-on, rotating disc galaxy, a small flattened galaxy and an irregular dwarf galaxy.

(iv) There are no signs of strong interactions in the form of tidal streams or H I clouds between the galaxies, down to a H I column density of $N_{HI} = 1.4 \times 10^{20}$ cm$^{-2}$. Similarly, there is no evidence of a triggered starburst or recent quenching in any of the galaxies as measured by the H I fraction, SSFR and SFE.

(v) The asymmetric optical morphology of IC 5201, its warped H I disc and the irregular morphology of AM 2220−460 are best explained by a close fly-by of AM 2220−460 with IC 5201 within the past $\sim 1$ Gyr. The IC 5201 orbital timescale with the fly-by timescale of AM 2220−460 are consistent with this scenario.

(vi) We have independently detected and verified the H I properties (flux, mass, $v_{sys}$, $w_{20}$) of ESO 289-G05, compared to paper II. This is important for the verification of ASKAP, ASKAPsoft and SoFiA, as slightly different observations, imaging techniques and significantly different RMS values exist between the images.

(vii) The new H I detection of GALEXASC J22 shows that its $v_{sys}$ is the same as the NGC 7232/3 triplet. Therefore, GALEXASC J22 is likely to be a new member of the NGC 7232/3 group, although there is no evidence of interaction with the triplet as of yet.

Overall, we have shown how ASKAP images the H I of a well resolved, extended, bright spiral galaxy. The greatest advantage of ASKAP over existing interferometers is its instantaneous wide-field imaging. This work shows the potential of resolving the H I in the disc as well as detecting and identifying close H I rich satellite galaxies on large survey scales (e.g. WALLABY).

# ACKNOWLEDGEMENTS

The Australian SKA Pathfinder (ASKAP) is part of the Australia Telescope National Facility (ATNF) which is managed by CSIRO. Operation of ASKAP is funded by the Australian Government with support from the National Collaborative Research Infrastructure Strategy (NCRIS). ASKAP uses the resources of the Pawsey Supercomputing Centre.

Establishment of ASKAP, the Murchison Radio-astronomy Observatory (MRO) and the Pawsey Supercomputing Centre are initiatives of the Australian Government, with support from the Government of Western Australia and the Science and Industry Endowment Fund. We acknowledge the Wajarri Yamatji as the traditional owners of the Observatory site. This work was supported by resources provided by the Pawsey Supercomputing Centre, including computational resources provided by the Australian Government under the National Computational Merit Allocation Scheme (project JA3).

The Parkes telescope is part of the Australia Telescope National Facility which is funded by the Australian Government for operation as a National Facility managed by CSIRO.

This work is based in part on observations made with the Galaxy Evolution Explorer (GALEX). *GALEX* is a NASA Small Explorer, whose mission was developed in co-operation with the Centre National d'Études Spatiales (CNES) of France and the Korean Ministry of Science and Technology. This work was supported by resources provided by the Pawsey Supercomputing Centre with funding from the Australian Government and the Government of Western Australia.

This project used public archival data from the Dark Energy Survey (DES). Funding for the DES Projects has been provided by the U.S. Department of Energy, the U.S. National Science Foundation, the Ministry of Science and Education of Spain, the Science and Technology FacilitiesCouncil of the United Kingdom, the Higher Education Funding Council for England, the National Center for Supercomputing Applications at the University of Illinois at Urbana-Champaign, the Kavli Institute of Cosmological Physics at the University of Chicago, the Center for Cosmology and Astro-Particle Physics at the Ohio State University, the Mitchell Institute for Fundamental Physics and Astronomy at Texas A&M University, Financiadora de Estudos e Projetos, Fundação Carlos Chagas Filho de Amparo à Pesquisa do Estado do Rio de Janeiro, Conselho Nacional de Desenvolvimento Científico e Tecnológico and the Ministério da Ciência, Tecnologia e Inovação, the Deutsche Forschungsgemeinschaft, and the Collaborating Institutions in the Dark Energy Survey.

The Collaborating Institutions are Argonne National Laboratory, the University of California at Santa Cruz, the University of Cambridge, Centro de Investigaciones Energéticas, Medioambientales y Tecnológicas-Madrid, the University of Chicago, University College London, the DES-Brazil Consortium, the University of Edinburgh, the Eidgenössische Technische Hochschule (ETH) Zürich, Fermi National Accelerator Laboratory, the University of Illinois at Urbana-Champaign, the Institut de Ciències de l'Espai (IEEC/CSIC), the Institut de Física d'Altes Energies, Lawrence Berkeley National Laboratory, the Ludwig-Maximilians Universität München and the associated Excellence Cluster Universe, the University of Michigan, the National Optical Astronomy Observatory, the University of Nottingham, The Ohio State University, the OzDES Membership Consortium, the University of Pennsylvania, the University of Portsmouth, SLAC National Accelerator Laboratory, Stanford University, the University of Sussex, and Texas A&M University.

Based in part on observations at Cerro Tololo Inter-American Observatory, National Optical Astronomy Observatory, which is operated by the Association of Universities for Research in Astronomy (AURA) under a cooperative agreement with the National Science Foundation.

DK thanks Kelly Hess for the tutorial outlining the steps used to make the rendered 3D image. PK is in parts supported by BMBF project 05A17PC2. We thank the referee and scientific editor for useful feedback that improved the quality of this manuscript.

This project has received funding from the European Research Council (ERC) under the European Union's Horizon 2020 research and innovation programme (grant agreement no. 679627).

This research was supported by the Australian Research Council Centre of Excellence for All Sky Astrophysics in 3 Dimensions (ASTRO 3D), through project number CE170100013

This paper has been typeset from a TEX/LATEX file prepared by the author.